\documentclass[prb,twocolumn,showpacs]{revtex4} 
\usepackage{graphicx}
\begin{document} 
\title{Inelastic Interaction
Corrections and Universal Relations for Full Counting Statistics}
\author{J. Tobiska and Yu. V. Nazarov} 
\affiliation{Kavli Institute of NanoScience Delft,\\Delft University of Technology,\\
Lorentzweg 1, 2628 CJ Delft, The Netherlands } 
\date{\today}
\pacs{73.23.-b, 72.10.-d, 72.70.+m}
\keywords{Full Counting Statistics,interaction correction,current,noise,third cumulant}
\begin{abstract} 
We analyze in detail the interaction correction
to Full Counting Statistics (FCS) of electron transfer in 
a quantum contact originating 
from the electromagnetic environment surrounding the contact.
The  correction can be presented as a sum of two terms, 
corresponding to elastic/inelastic
electron transfer. Here we primarily 
focus on the inelastic correction.

For our analysis, it is important to understand more general --- 
universal  --- relations imposed on FCS only by quantum mechanics and
statistics with no regard for a concrete realization 
of a contact. So we derive and analyze these relations. 
We reveal that for FCS the universal relations can be presented in
a form of detailed balance. We also present several useful formulas
for the cumulants.

To facilitate the experimental observation of the effect,
we evaluate cumulants of FCS at finite voltage and temperature.
Several analytical results obtained 
are supplemented by numerical calculations for the first
three cumulants at various transmission eigenvalues. 
\end{abstract} 
\maketitle

\section{Introduction}
 Within the 
last years technological advancements
have enabled the fabrication of sufficiently small
(nanometer)
solid state structures where electrons traverse
the system as coherent quantum waves.
The electron transport in such a quantum contact 
can be described with the scattering approach pioneered
by Landauer and B{\"u}ttiker \cite{LandauerButtiker1,LandauerButtiker2,LandauerButtiker3}.
Within this approach, the contact is completely
characterized by the set of transmission eigenvalues $\{T_n\}$,
$0 < T_n < 1$,
regardless its concrete structure. 
The (differential) conductance is given by the Landauer-B{\"u}ttiker formula
$G=G_\mathrm{Q}\sum_n T_n, G_Q =2e^2/h$.
The same transmission eigenvalues determine superconducting 
and noise properties of the structure
\cite{generalTn1,generalTn2}.
Break junction experiments 
\cite{breakjunctions} provide excellent examples of tuning and
experimental characterization of $T_n$'s in concrete quantum contacts.
In the scattering approach, the electron-electron interaction
{\it inside} the contact 
is commonly disregarded. There is a good reason for that, eventually
the same as for electrons in bulk metallic solids.
Close to the Fermi energy, the only effect of interaction is to make
a quasi-particle from an electron. These quasi-particles do not interact.
This means that any contact at sufficiently low energies
can be described within a non-interacting scattering approach.

This however presumes an ideal voltage bias 
of the contact: the electrons are injected and absorbed
by reservoirs kept at a certain voltage.
This assumption is too ideal: in reality, the contact is
embedded in a macroscopic electric circuit, and this 
{\it electromagnetic environment} produces
voltage fluctuations on the contact. 
The electrons traversing the contact can emit/absorb energy 
to/from the environment and interact by means of exchange of
photons that are present in the environment. The
interaction due to the environment can not be disregarded at low
energies and therefore becomes the most important interaction 
at low temperature and voltage.
The environment is completely characterized by a 
frequency dependent impedance $Z_\omega$, in series with the quantum contact.

The influence of the environment on electron transport has been 
studied in detail for tunnel junctions where all $T_n \ll 1$
\cite{IngoldNazarov}. The tunnel rates in the presence of an environment 
can be evaluated for arbitrary impedance.
For sufficiently large environmental impedances
 $ZG_Q/2 \equiv z\gg 1$  the interaction effects are large 
 and the tunnel rates are strongly suppressed below a
certain energy. This is termed Coulomb blockade of tunneling 
in a single junction \cite{AverinLikharev}. 
The opposite case of small impedance $z \ll 1$ is more realistic.
In this case, the environment provides an interaction correction $\simeq z$
to the rates. This correction can be experimentally identified
from its specific voltage and temperature dependence: 
it is seen as so-called zero-bias anomaly.
\cite{Yuli_anomaly, KuzminLikharevOld}

At arbitrary transmissions, the influence of the environment
is more complicated and one cannot evaluate it for an arbitrary impedance
(a progress in this direction has been reported in \cite{Safi}).
Still one can investigate the interaction correction $\simeq z$ to 
the contact conductance,
both theoretically and experimentally.
 It has been demonstrated~\cite{Zaikin,yeyati0bias}
that this correction is related
to the second moment of current fluctuations: noise. 
The correction is proportional to shot noise $\sim T_n(1-T_n)$,
and disappears at perfect transmissions $T_n =1$.
This prediction has been experimentally confirmed.\cite{QPC-exp}

The environment influences not only the average current, but the whole
statistics of electron transfers in the contact, the Full Counting
Statistics (FCS). The theory of FCS for a quantum contact
within the scattering formalism has been developed
in~\cite{jmphyslevles}. Later, it has been 
successfully applied to a variety of
systems. The FCS is the  statistics
of current measurements 
over a given time interval $\tau$, $\tau$ being much
bigger than the typical time between electron transfers.
It gives the probability $P_\tau(N)$ for $N$ electrons
to be transferred during this time interval. 
It is convenient to work with the generating function 
defined as
\begin{equation}
F(V,\chi) = \sum_N P_\tau(N) e^{i\chi N}.
\end{equation}
The parameter $\chi$ is frequently called counting field
since if one implements the Keldysh formalism for FCS \cite{yuliann}
$\chi$ enters the formulation as a field conjugated to
electric current.
Derivatives of $\ln F(\chi)$ with respect to $\chi$
give an infinite set of cumulants of charge transferred, 
where the first two are related to average current and current noise. 

The environment influences the FCS in two distinct ways.
A classical effect is governed by the impedance at low frequencies
$\hbar\omega \ll \max(eV,k_\mathrm{B} T)$ and scales like 
$ZG$, $G$ being the total conductance of the contact. For the two first
cumulants --- average current and noise --- this effect is nothing
but voltage division between the contact and the external impedance.
The effect is not that trivial for higher cumulants 
\cite{kindermcurvoltbias}, and the pioneering measurement
of the third
cumulant~\cite{thrdprl} was shown to be affected by this environmental
effect. However, this effect can be made arbitrarily small by
a proper design of the low-frequency impedance to assure $ZG \ll 1$.

A more interesting effect comes from the impedance 
at frequencies $\hbar\omega \gtrsim \max(eV,k_\mathrm{B} T)$
and scales as $ZG_Q$. This is the interaction correction
discussed which eventually leads to Coulomb blockade effects
at higher impedance. It has been shown in \cite{kindermann:136802}
that the correction can be separated into inelastic and elastic 
parts, the latter presents a renormalization of elastic scattering
properties of the contact by interaction. It is feasible to
observe the interaction correction to FCS in experiments, 
for instance, with well-characterizable break junctions.
Such experiments would be certainly possible for the first three cumulants,
and the developments in the field \cite{FCS-measurement1,FCS-measurement2,FCS-measurement3,FCS-measurement4} 
suggest that the higher
cumulants can be accessed with proper measurement techniques as well.
Apparently, the correction will include both the elastic 
and inelastic part. To provide theoretical support 
for these experiments is the main motivation and goal
of the present work which concentrates on the inelastic part of the interaction
correction.

In Section \ref{sec:action} we derive the interaction correction to 
the FCS
in first order perturbation in $z$ starting from the general form 
of the system-environment
Keldysh action. The result allows for the
identification of an elastic and inelastic contribution.
In Section \ref{sec:interpretation} we explain how one can re-interpret
this result in terms of correlations of elementary events
of charge transfer.
We present concrete analytical results in the next Section, particularly
for vanishing temperature.

While investigating the interaction correction in the limit of small voltages,
we have found sets of simple relations for cumulants. Further
analysis has shown that these relations are not specific for the
setup considered and hold for any conductor regardless
its properties and presence or absence of interaction. 
In fact, these universal relations provide 
the generalization of the fluctuation-dissipation
theorem~\cite{FDT} for FCS. We discuss these relations
at length in Section \ref{sec:universal}. The derivation is provided
in the Appendix.

In the last Section we present our numerical results
for arbitrary temperature and voltage. We 
have studied the experimentally interesting case
of an RC-environment and plot the correction to the first three cumulants
for conductors of different transmissions and a diffusive conductor
versus $eV/k_\mathrm{B}T$ as obtained from a numerical evaluation of the integrals.
In all cases we observe a crossover at $eV\sim k_\mathrm{B}T$ which is related to a transition from thermal to shot noise behavior.

\section{Action}
\label{sec:action}

The system we consider consists of a quantum conductor which can be
described by the set of its transmission probabilities,
$T_n$, and a frequency dependent environmental impedance,
$z_\omega$ in series. The voltage drop 
over the whole conductor-environment system is fixed. 
However, the voltage in the node between the contact and the
impedance can fluctuate. For instance, an
 electron transfered will momentarily charge the node 
creating a voltage pulse $\propto z$ that persists for
some time and may influence
further electron transfer. 
Thus there will be a fluctuating voltage in the node
depending both on the probabilistic nature of the
electron transfer in the quantum conductor 
as well as on the impedance of the environment. We 
study the corrections to electron transport due to these fluctuations.

It is convenient to work with the phase rather than voltage,
which is defined as $\phi=\int dt eV(t)/\hbar$.
Since we study a quantum mechanical system, we have to describe
quantum fluctuations of this quantity. This is most conveniently
presented in the language of Keldysh action that
expresses physical quantities in the form of 
 a path integral over the fluctuating phase on the Keldysh contour.
Since the Keldysh contour consists of two parts, the
integration proceeds over two sets of variables $\phi^{\pm}(t)$
corresponding to these parts.

The Keldysh action approach to mesoscopic quantum circuits
has been pioneered in~\cite{schoen} and has been extended 
to cover FCS and arbitrary quantum contacts in~\cite{kindermcurvoltbias,kindermannnazarov}.
As usual in the theory of FCS, it is the 
generating function $F(\chi)$ which is presented 
as a path integral over the fluctuating phase. 
The action in the path integral 
is a sum of the actions describing the constituent parts of the circuit: 
the conductor action, $S_\mathrm{c}$, and the environment action, 
$S_\mathrm{env}$,
\begin{eqnarray}
	F(\chi)&=&\int d\phi^+d\phi^-\exp\{-i S[\phi^+,\phi^-]\};\nonumber \\
	S[\phi^+,\phi^-] &=&S_\mathrm{c}[\phi^+,\phi^-]+ \nonumber\\
+ &S_\mathrm{env}&[\Phi+\chi/2-\phi^+,\Phi-\chi/2-\phi^-] .
\label{cgf}
\end{eqnarray}
We use superscripts $\pm$ to denote the phases 
at different parts of the contour
and use traditional 
notations $\varphi$ ($\chi$) for their half-sum (difference). 
Current, noise and
higher moments of FCS follow as $\partial \ln F/\partial \chi|_{\chi=0},\partial^2
\ln F/\partial\chi^2|_{\chi=0},\cdots$. 
For a linear environment the action $S_\mathrm{env}$
is a bilinear function of the phases
depending on the impedance and temperature $T$ only,
\begin{eqnarray}
	S_\mathrm{env}=\frac{1}{2\pi}\int_0^\tau dt\int_0^\tau dt' \nonumber \\
	\left[ (\phi^+(t),\phi^-(t))A(t-t')\left({\phi^+(t')\atop\phi^-(t')}\right)\right],\label{envactn}
\end{eqnarray}
where the coefficient matrix, $A$, depends solely on frequency and temperature:
\begin{equation}
	A(\omega)=\left(\begin{array}{cc}
		-i\omega[z^{-1}_\omega+2N_\omega \Re z^{-1}_\omega]&2i\omega N_\omega\Re z^{-1}_\omega\\
		-[A^{+-}(-\omega)]^*&-[A^{++}(-\omega)]^*
	\end{array}\right).\label{amtrx}
\end{equation}
$N_\omega=\{\exp[\hbar\omega/k_\mathrm{B}T]-1\}^{-1}$ 
is the Bose-Einstein distribution function. 

The most concise way to write the conductor 
action is in terms of Keldysh-Green 
functions $\hat{L},\hat{R}$ of the 
left/right reservoir~\cite{yuliann}:
\begin{equation}
	S_\mathrm{c}=\frac{i}{2}\sum_n\mathrm{Tr}\ln\left[1+\frac{T_n}{4}(\{\hat{L},\hat{R}\}-2)\right].\label{condactn}
\end{equation}
The trace is over Keldysh space and energy/frequencies.
In equation~(\ref{condactn}) 
the fluctuating phase in the node
enters 
in the form of a gauge transform 
in one of the reservoirs as
$\hat{L}=e^{i\phi}G^\mathrm{res}e^{-i\phi}$, 
$\hat{R}=G^\mathrm{res}$, where $\phi=\varphi+\chi/2\tau_z$. 
The equilibrium Keldysh-Green function depends on time difference
(or energy) only and reads in a usual way
\begin{equation}
	G^\mathrm{res}=\left(\begin{array}{cc}
		1-2f(\varepsilon)&2f(\varepsilon)\\
		2(1-f(\varepsilon)&2f(\varepsilon)-1
	\end{array}\right),\label{resG}
\end{equation}
with $f$ being the Fermi distribution function of the corresponding lead. 
Equations (\ref{cgf},\ref{envactn},\ref{amtrx},\ref{condactn},\ref{resG}) 
define our model and all we have to calculate is 
the cumulant generating function $F$ for a 
given environment $z_\omega$ and conductor $T_n$. 
However in the general case this is a formidable task.
A natural way 
to proceed is to assume $z\ll 1$ 
and thus to treat the effect of the environment 
as a perturbation. 
In zeroth order (no environment), 
the phases are obviously related to the applied voltage and do not fluctuate: 
$\varphi=eVt$.

Putting this in equation (\ref{condactn}) and taking the trace gives the
well known expression for the generating function of a mesoscopic
conductor in terms of its transmission eigenvalues~\cite{jmphyslevles}
\begin{eqnarray}
	\ln F^{(0)}(\chi)\equiv \frac{\tau}{\hbar}S^{(0)}=\frac{\tau}{\hbar}\sum_n\int\frac{d\varepsilon}{2\pi}
	\nonumber\\
	\ln\left\{1+T_n(e^{i\chi}-1)f_l(1-f_r)+(e^{-i\chi}-1)f_r(1-f_l)\right\}
	\label{eq:elastic}
\end{eqnarray}
Here and in 
the following indices 'l,r' refer to the 
left/right lead. We assign the voltage to the left lead, so that 
$f_l(\varepsilon+V)=f_r(\varepsilon)=f(\varepsilon)$.
	
The first order correction is proportional to the fluctuations of
$\phi$, which are small $\propto z$. 
Expanding the logarithm in equation (\ref{condactn}) to second
order in $\phi$ gives a second order contribution
\begin{eqnarray}
	-i\delta S_\mathrm{c}^{(2)}&=&\frac{T_n}{8}\mathrm{Tr}\{D A^{(2)}\}-\frac{1}{4}\left(\frac{T_n}{4}\right)^2\mathrm{Tr}\{DA^{(1)}DA^{(1)}\}\\
	&=&\frac{T_n}{8}D\mathrm{Tr}\{A^{(2)}\}-\frac{1}{4}\left(\frac{T_n}{4}\right)^2DD^+\mathrm{Tr}\{A^{(1)}A^{(1)}\}
\end{eqnarray}
where the following relations hold under the trace
\begin{eqnarray}
A^{(1)}A^{(1)}=&&4\phi^2-4\phi L^+\phi L-2\phi L^+R^+\phi LR \nonumber \\
&&-2\phi R^+L^+\phi RL+4\phi L^+R^+L^+\phi R,\\
A^{(2)}=&&\phi L^+\phi R+\phi R^+\phi L-\phi^2 (RL+LR)\label{eq:a2exp}
\end{eqnarray}
and
\begin{eqnarray}
	D^{-1}=1+T_n[(e^{i\chi}-1)f_l(1-f_r)+ \nonumber \\
	+(e^{-i\chi}-1)f_r(1-f_l)].
\end{eqnarray}
All quantities with superscript 
'$+$' are taken at energy $\varepsilon+\omega$ and 
integration over energy and frequency is implied. For convenience we omitted the explicit dependence on $\varepsilon,\omega$. It is easily found from the definition of the trace. The first term in equation (\ref{eq:a2exp}) for instance would read $T_n/8\mathrm{Tr}\int d\varepsilon d\omega D_\varepsilon\phi_{-\omega}L_{\varepsilon+\omega}\phi_\omega R_\varepsilon$, where the trace is understood over Keldysh indices.

These terms are quadratic in phase, 
and by virtue of the path integral in equation (\ref{cgf}) 
are to be replaced 
with their averages given by the environmental action. 
These averages read
\begin{equation}
\langle\left(\begin{array}{cc}
	|\varphi|^2&\varphi\chi^*\\
	\varphi^*\chi&|\chi|^2
\end{array}\right)\rangle\to
	\left(\begin{array}{cc}
	(2 N_\omega+1)\frac{\Re z_\omega}{\omega}&\frac{z_\omega}{\omega}\\
	-\frac{z^*_\omega}{\omega}&0
\end{array}\right).
\end{equation}

After some ordering of terms,
the first order correction to the cumulant
generating function can be presented as
\begin{eqnarray}
	(\ln F(\chi))^{(1)}=\frac{\tau}{\hbar}\int_0^\infty d\omega\frac{\Re z_\omega}{\omega}\left[(2N_\omega+1)S_\mathrm{el}^{(1)}(\chi)+\right. \nonumber \\
	\left. N_\omega S_\mathrm{in}^{(1)}(\omega,\chi)+ (N_\omega +1) S_\mathrm{in}^{(1)}(-\omega,\chi)\right].\label{eq:fstorder}
\end{eqnarray}
We note that there are three different 
terms which can be identified as being due 
to elastic electron transfer, 
and inelastic transfer with either absorption (positve $\omega$)
or emission (negative $\omega$) 
of energy $\hbar\omega$ from/to the environment. 
Explicitly, in terms of filling factors these terms read
\begin{eqnarray}
&	S_\mathrm{el}=&-2\sum_n T_n(1-T_n)\frac{\partial S^{(0)}}{\partial T_n}\\
&S_\mathrm{in}=&\sum_n\int\frac{d\varepsilon}{2\pi}\left\{DD^+[T_n(f_l-f_l^+)\right.\nonumber\\
&&	\left.+2T_n(e^{i\chi}-1)f_l(1-f_r^+)\right. \nonumber\\
&&	\left.+4T_n^2(\cos\chi-1)f_l(1-f_l^+)(f_r^+-f_r)]\right.\nonumber\\
&&	\left.+(1-D)(1-T_n-D^+)\right\}+
	\left\{\begin{array}{c}l\leftrightarrow r\\\chi\leftrightarrow -\chi
	\end{array}\right\} \label{eq:inelastic}
\end{eqnarray}
Since the expression is symmetric with respect to 
exchange $l \leftrightarrow r$ of the leads
and simultaneous change of the sign of the counting field,
the cumulants are either 
 even or odd functions of the voltage applied. 
Following \cite{kindermann:136802} we present 
the elastic part of the correction as a change $\propto z$ 
of transmission eigenvalues. It is clear that the main contribution to the inelastic part of the
integral in equation (\ref{eq:fstorder}) comes from small frequencies less
than the maximum of voltage and temperature. Elastic processes do not
have this restriction and $S_\mathrm{in}$ contributes to the integral at
all frequencies.

\section{Interpretation}\label{sec:interpretation} 
The advantage of the FCS approach to quantum transport 
is that in many cases the FCS expression can be re-interpreted
in terms of elementary events thus providing some insight 
into the relevant transport properties. A well-known example
of such interpretation is provided by the non-interacting 
Levitov formula \cite{jmphyslevles}. At vanishing temperature,
it allows to present the statistics as a superposition
of $\tau eV/\hbar$ elementary "games" in each transport channel
$n$ , each "game"
resulting in either transmission (with chance $T_n$) or
reflection (with chance $1-T_n$) of an electron coming to the contact.
A recent example of such a re-interpretation concerns spin statistics
\cite{AntonioYuli}. In this Section, we show how to interpret
the inelastic interaction correction given by Eq. \ref{eq:inelastic}.
As we will see, the interpretation is probably too complicated
to be constructive. Still it gives some insight to the form of
the expression, in particular, the presence of denominators $D,D^{+}$.

To give an interpretation of the form of
$S_\mathrm{in}$ let us go back to relation (\ref{eq:elastic}) that
holds for non-interacting electrons. In the limit of large measurement 
time $\tau$,
$\hbar/\tau\ll \max(k_\mathrm{B}T,eV)$, we can discretize the
integration over energies. The general structure of the generating
function~\cite{levitov:230} is then

\begin{equation}
	F^{(0)}(\chi)=\prod_E \lambda(E), \label{eq:fstrct}
\end{equation} 

with $\lambda(E)=D^{-1}$. 
The product is taken 
over discretized energies as well as over transport channels. 
The right hand side of
equation (\ref{eq:fstrct}) is a product,
so each term in this product can be regarded as an independent 
process ("game").
The concrete form of $\lambda(E)$ suggests that
 three distinct outcomes of each process are possible: 
 (i) electrons transmitted
from the left reservoir to the right with probability $P_+=T_n
f_l(1-f_r)$, (ii) transmission from right to left with
$P_-=T_n f_r(1-f_l)$, (iii) no transmission
($P_0=1-P_+-P_-$). Indeed, 
the generating function for each process is then
$\lambda=\sum_\alpha P_\alpha X_\alpha$ with $X_\alpha=e^{i\alpha\chi}$. Electrons at different energy (and/or channel) are uncorrelated
since the complete generating function
factorizes in terms of $\lambda(E)$.

How can this picture change if one introduces electron-electron
interaction via  the environment? It is clear that
interaction will bring about all kinds of correlations between electrons
at different energies 
and the simple picture presented above is not true anymore. The major
change is, that the factorization in uncorrelated elementary events does
not hold. Presumably, the generating function
of an elementary event $\Lambda$ will depend on
many different energies. The relation (\ref{eq:inelastic})
suggests that in lowest order in $z$  it
depends on two energies only, $\Lambda(E,E')$,
 where $|E-E'|=\hbar\omega$ is the energy
of an absorbed/emitted photon. 
With this accuracy, the cumulant generating
function can be expressed as a product
over pairs of energies
\begin{equation}
F^{(0)}(\chi)+ F^{(1)}(\chi)=\prod_{E,E'}\Lambda(E,E').
\end{equation}
Without interactions,
\begin{equation}
\Lambda(E,E') =\left\{\begin{array}{cc} \lambda(E) & {\rm if}\ E=E' \\
1 & {\rm if}\ E \neq E' \end{array} \right.
\end{equation}
so that electrons with different energy are uncorrelated. 
If $\delta \Lambda(E,E')$ is the interaction correction to
$\Lambda$, the change of the cumulant generating function reads
\begin{equation}
\left(\ln F(\chi)\right)^{(1)} = \sum_{E,E'} \frac{\delta \Lambda(E,E')}
{\lambda(E)\lambda(E')}
\end{equation}
This explains already the presence of denominators in (\ref{eq:inelastic}).
In addition, we conclude from  (\ref{eq:inelastic})  that contains 
a single sum over transport channels, that the elementary
events do not involve electrons in different channels,
even though they involve electrons at 
different energies. 
This is probably valid only for the first order correction.

To proceed, let us note that
the correction  $\delta \Lambda(E,E')$ 
consists of terms to be divided into three classes. 
Firstly, there will be terms presenting {\em new-events},
$\delta\Lambda_\mathrm{new}(E,E')$ not taking place
for non-interacting electrons. An example is an electron transfer from
the left to the right with photon emission.
As we can assert from (\ref{eq:inelastic})
it comes with probability $\propto T_n f_l(E)(1-f_r(E'))(1+N_{E-E'})$.
Another example is a
two-particle process 
consisting of elastic electron transfer at energy $E$ accompanied
by inelastic transfer, its probability being proportional to
$T_n^2 f_l(E)f_r(E)(1-f_l(E'))(1-f_r(E))(1+N_{E-E'})$.
Secondly, since the probabilities of 
elementary {\em old-events} are modified by interaction,
there will be terms depending on a single energy only,
those can be seen as the modification of $\lambda(E)$,
$\lambda(E) \rightarrow \lambda(E)+\delta\Lambda_\mathrm{old}(E,E)$. 
They are incorporated into the elastic part of the correction.
Finally, the environment will introduce {\em
correlations} among pairs of old-events, represented 
by $\delta\Lambda_\mathrm{corr}(E,E')$.  For instance,
the correlation between left-right transfer at energy $E$ and
right-left transfer at energy $E'$ will come with a factor 
$T^2_n f_l(E)(1-f_r(E)) f_r(E')(1-f_l(E'))$.

These three contributions 
simply add up in the correction 
to the generating function, 
\begin{eqnarray}
\left(\ln F(\chi)\right)^{(1)}=
\sum_E\frac{\delta\Lambda_\mathrm{old}(E,E)
}{\lambda(E)} \cr
+2 \sum_{E>E'}
\frac{\delta\Lambda_\mathrm{new}(E,E')+\delta\Lambda_\mathrm{corr}(E,E')}
{\lambda(E)\lambda(E')}.
\end{eqnarray}
One recognizes this structure 
in the relations for $S_\mathrm{el}$ (first term)
and for $S_\mathrm{in}$ (second term).
In principle, in this way one can recover the correction 
to the generating function of an elementary event $\delta \Lambda$
and find the (corrections to) probabilities of all possible 
outcomes, new ones as well as old ones. This gives a re-interpretation
of the correction: any term of equation (\ref{eq:inelastic}) 
is assigned to a term of
one of the three classes discussed.

However, the procedure is cumbersome and hardly practical
because of the large number of possible 
processes and outcomes. 
For a  two-electron
process, each incoming electron can be in
one of four possible states   (coming from the left or the right 
at $E$ or  $E'$), the same for outgoing electrons. This gives in total
$2^8$ terms: it looks like a somewhat lengthy
interpretation of a relatively compact 
equation (\ref{eq:inelastic}). This prevented us from accomplishing this
program explicitly. We are satisfied with the fact that
the combinations of electron and photon filling factors make sense
for the terms contributing to $\delta \Lambda_{{\rm new},{\rm corr}}$.
The picture and the interpretation are expected
to become even more involved for the corrections of higher orders in $z$.

\section{Analytical Results}\label{sec:analytical}

At vanishing temperature we can easily perform the integration over
$\varepsilon$ in equation~(\ref{eq:fstorder}).
The full correction to FCS then reads
\begin{eqnarray}
\left(\ln F(\chi)\right)^{(1)} = 
\frac{\tau}{\hbar} eV S_\mathrm{el} 
\int_{eV/\hbar}^\infty d\omega\frac{\Re z_\omega}{\omega};\\
S_\mathrm{el} = -2 \sum_n T_n(1-T_n) \frac{e^{i\chi}-1}{1+T_n(e^{i\chi}-1)} . \nonumber
\end{eqnarray}
It has a simple and interesting structure revealing the 
relationship between the elastic and inelastic part of the correction.
If we took the elastic part only, by virtue of
(\ref{eq:elastic}) we would obtain a similar expression. The difference
is that the integration over $\omega$ would start at zero.
Therefore, the inelastic part of the correction precisely cancels 
the modification of the elastic transmission for photon energies
in the interval $0<\hbar\omega<eV$. It is indeed expected
from general reasoning \cite{kindermann:136802} 
that the low-energy divergences
present in the elastic correction are cut off at energies $\simeq eV$.
Our somewhat unexpected result is that at vanishing temperatures
this cut-off is sharp and clear. In agreement with expectations, 
the inelastic $S_\mathrm{in}$ only contributes at 
frequencies $0<\hbar\omega<eV$ reflecting 
the fact that the only energy source 
for inelastic processes is given by the voltage.

The correction to
the $m$-th cumulant $S^{(m)}$ is given by
derivatives of the above relation, we find:
\begin{eqnarray}
	\delta S^{(m)}&=& -2\frac{\tau}{\hbar}eV \sum_n
T_n(1-T_n)\nonumber\\
&&\left.\frac{d^m}{d(i\chi)^m}\frac{e^{i\chi}-1}{1+T_n (e^{i\chi}-1)}\right|_{\chi=0} \int_{eV/\hbar}^\infty d\omega\frac{\Re z_\omega}{\omega}\label{zerotcum}
\label{eq:cumulant-big-voltage}
\end{eqnarray}
Importantly, the correction 
is proportional to the $m+1$-th cumulant for the non-interacting case,
$\delta S^{(m)} \propto  S^{(m+1)}$. This generalizes the
results \cite{Zaikin,yeyati0bias} for the average current.
The environment enters  the corrections as an integral 
over the impedance and affects every cumulant in the same way. 

A common and  experimentally interesting model for a
frequency-dependent impedance is that of an RC-environment,
$z_\omega=z(1+i\omega/\omega_\mathrm{c})^{-1}$. 
The impedance is cut at $\omega_\mathrm{c}=1/(RC)$ and approaches a constant value of $z$ at $\omega \ll \omega_\mathrm{c}$.
The integral governing the correction evaluates to
\begin{equation} 
\int_{eV/\hbar}^\infty d\omega\frac{\Re
z_\omega}{\omega}=
\ln\sqrt{1+\frac{\hbar^2\omega_\mathrm{c}^2}{e^2V^2}}\approx\ln\frac{\hbar\omega_\mathrm{c}}{eV},~
\mathrm{if}~\hbar\omega_\mathrm{c}\gg eV \label{zerotum} \end{equation} 
That is, it diverges logarithmically 
at sufficiently low voltages $eV \ll \hbar \omega_\mathrm{c}$. 
This is the
well known zero bias anomaly. As has been shown it holds for any
cumulant. Different cumulants have the same functional dependence on
voltage and can be scaled by the prefactor of equation (\ref{zerotum})
which depends only on the transmission probabilities. 

Even if $z \ll 1$ the correction $\propto z \ln(\hbar\omega_\mathrm{c}/eV)$ 
can become big at sufficiently small voltages, $eV/\hbar\omega_\mathrm{c} \simeq e^{-1/2z}$.
It has been shown in ~\cite{kindermann:136802} that in this case one has 
to implement the renormalization procedure
neglecting the inelastic part. The elastic correction 
can be consolidated in the energy dependent renormalization of 
transmission eigenvalues given by
\begin{equation} 
	\frac{dT_n(E)}{d\log
E}=2zT_n(E)(1-T_n(E)) .
\end{equation}

We do not consider this here. Rather, we expect that 
finite temperature
will lead to a rounding off of the singularity 
at small voltages in the same way
as for the current correction.

The equation (\ref{eq:fstorder}) is too complex at finite temperatures
so that it is hard and
non-instructive to perform the integration over energies $\varepsilon$. 
However, the
correction to any cumulant of finite order 
derived from the generating function is an integral
over a finite polynomial of Fermi functions.
This integration can easily be done for
arbitrary temperature and voltage. The analytical
formulas obtained in this way are too long to be of any use
except numerical evaluation.
The correction to any 
cumulant is proportional to $T_n(1-T_n)$, thereby 
vanishing at perfect and vanishing transmission.

For the correction to the average current we find
\begin{eqnarray}
	&\delta I =& -2e\sum_n T_n(1-T_n) \nonumber\\
	&&\int_0^\infty d\omega \frac{\Re z_\omega}{\omega}\frac{\omega \sinh \frac{eV}{k_\mathrm{B}T}-\frac{eV}{\hbar}\sinh\frac{\hbar\omega}{k_\mathrm{B}T}}{\cosh \frac{eV}{k_\mathrm{B}T}-\cosh\frac{\hbar\omega}{k_\mathrm{B}T}}\label{eq:currexactlong}\\
	&=&-e\int_0^\infty d\omega\frac{\Re z_\omega}{\omega}\langle|\Delta I|^2_\omega\rangle ,
	\label{eq:currexact}
\end{eqnarray}
where $\langle|\Delta I|^2_\omega\rangle$ is the finite-frequency current correlator without environment~\cite{yeyati0bias,shotnoise}. The correction to the current is thus related to the noise in the absence of an environment.

The correction to the $m$-th cumulant is an $m+1$-th order polynomial
in $T_n$. The term linear in $T_n$ 
has the same functional 
dependence as that in (\ref{eq:currexact}). 
The reason for this is that in the limit of small $T_n$
the Full Counting Statistics is that of a tunnel junction:
electron transfers are rare and consequently independent.
We get a superposition of two Poissonian statistics for
electrons tunneling from the left to the right and from the 
right to the left expressed as
\begin{equation}
\ln F(\chi) = \tau \left(
\Gamma_{LR}(V,T)(e^{i\chi} -1) + 
\Gamma_{RL}(V,T)(e^{-i\chi} -1) \right)
\end{equation}
$\Gamma_{LR,RL}$ being tunneling rates in these two directions.
The interaction in this limit modifies $\Gamma_{LR,RL}$,
this being the only effect on FCS.
We will see below that these two rates are related
by the detailed balance condition $\Gamma_{RL} =\Gamma_{LR} \exp(-eV/k_\mathrm{B}T)$.
From this it follows that in the tunneling limit
\begin{equation}
	\frac{e}{\tau}S^{(m)} = \left\{\begin{array}{ll}
		I&\mathrm{if}~m~\mathrm{odd}\\
		I \coth(\frac{eV}{2k_\mathrm{B}T})&\mathrm{if}~m~\mathrm{even}
	\end{array}\right.
\end{equation}
for any interactions. 

Analytical work in the limit of small voltages $eV \ll k_\mathrm{B}T$ 
gave us some relations between the cumulants. However,
we have recognized that these relations are not specific for the
interaction correction but are of general nature. That is why we discuss
them in the next Section.

\section{Universal relations for cumulants}\label{sec:universal}
The detailed investigation of the interaction corrections to FCS 
is hardly effective
without appreciation of universal relations for FCS cumulants 
that hold with no regard for interaction and/or concrete 
structure of the conductor. This is why in the course of this research
we had to understand the general constraints  imposed on
the FCS by laws of quantum mechanics and thermodynamics.

We show in the Appendix that this universal relation can be most
generally expressed in the following form
\begin{equation}
	F(V,\chi) = F(V,-\chi + ieV/k_\mathrm{B}T) .
\label{eq:UniversalRelation}
\end{equation}
A didactic representation of this relation can be obtained 
by recalling the definition of $F$ as generating function
of the probability distribution of a certain number $N$ of particle 
transfers,
\begin{equation}
F(V,\chi) = \sum_N P_N e^{i\chi N}.
\end{equation}
Applying (\ref{eq:UniversalRelation}), we observe that the probabilities
of opposite number of particles transferred are related by
\begin{equation}
	P_N(V) = P_{-N}(V) e^{eV N/k_\mathrm{B}T} . \label{eq:detailedbalance}
\end{equation}
This relation is well-known for independent tunnelling events (see, e.g. 
\cite{AverinLikharev,IngoldNazarov})
and is referred to as {\it detailed balance} condition.
We thus demonstrate that the detailed balance holds for any $N$ irrespective
of possible interactions and correlations in and beyond the conductor. 

Whatever didactic, the detailed balance condition is not easy to apply 
to cumulants.
We do this with the universal relation (\ref{eq:UniversalRelation}) 
that obviously holds for $\ln F$ as well. A series of
relations for cumulants is obtained by taking
derivatives of this relation with respect to voltage at $V=0$.
First of all, we just go to the equilibrium limit $V \rightarrow 0$
to obtain 
\begin{equation}
\ln F(\chi) = \ln F(-\chi),
\end{equation}
all even cumulants thus vanish at equilibrium.
This relation and {\it all subsequent relations} till the end
of this Section are valid only in the limit $V \rightarrow 0$.

Taking the first derivative with respect to voltage, we arrive at
\begin{equation}
	\frac{\partial }{\partial eV}\left(\ln F(\chi)-\ln F(-\chi)\right)
	=-\frac{i}{k_\mathrm{B}T}\frac{\partial \ln F(\chi)}{\partial \chi}.
\label{eq:basicprop}
\end{equation}
If we expand this in $\chi$, we obtain a series of equations 
that relate even cumulants with voltage derivatives of odd cumulants
\begin{equation}
	S^{(2n+2)} = \frac{2k_\mathrm{B}T}{e} \frac{\partial{S^{(2n+1)}}}{\partial V} 
\end{equation}
The first equation in this series is nothing but Johnson's noise
formula,
\begin{equation}
	e^2S^{(2)} = 2k_\mathrm{B}T\frac{\partial{I}}{\partial{V}},
\end{equation}
that relates zero-voltage conductance and equilibrium current noise.

The next series is obtained by taking the second derivative with
 respect to voltage and making use of (\ref{eq:basicprop}),
\begin{eqnarray}
\frac{\partial^2 }{\partial V^2}\left( \ln F(\chi)-\ln F(-\chi)\right) = \nonumber\\
\frac{-ie}{k_\mathrm{B}T} \frac{\partial^2 }{\partial V \partial \chi}\left( \ln F(\chi)+\ln F(-\chi)\right)
\end{eqnarray}
which is only practical if even(odd) cumulants are {\it not} even(odd) 
functions of voltage, that is, in the absence of electron-hole symmetry. 
Since our model is electron-hole symmetric, this relation is of no immediate
relevance. The first relation in the series reads
\begin{equation}
	\frac{\partial^2 I}{\partial V^2} = - \frac{e^2}{k_\mathrm{B}T}\frac{\partial S^{(2)}}{\partial V}.
\label{eq:series2}
\end{equation}
It relates d.c. current induced by low-frequency a.c. voltage
(rectification effect)
to low-frequency current noise proportional to d.c. voltage applied.  
This relation was discussed in detail in \cite{YuliOld} in the context of the
photovoltaic effect.

Taking the third derivative with respect to voltage
and making use of (\ref{eq:series2})
we obtain another series:
\begin{eqnarray}
\frac{\partial^3 }{\partial V^3}\left( \ln F(\chi)-\ln F(-\chi)\right) = \frac{ie^3}{(k_\mathrm{B}T)^3} \frac{\partial^3 \ln F(\chi)} {\partial \chi^3}\nonumber \\
-\frac{3ie}{k_\mathrm{B}T} 
\frac{\partial^3 }{\partial V^2 \partial \chi}
\left( \ln F(\chi)+\ln F(-\chi)\right)
\end{eqnarray}
The first relation in the series can be rewritten as 
\begin{equation}
2e^2 \frac{\partial^2 S^{(2)}}{\partial V^2} = \frac{1}{3}
\left( 2 k_\mathrm{B}T \frac{\partial^3 I}{\partial V^3} - 
\frac{e^4}{(k_\mathrm{B}T)^2} S^{(4)} \right) 
\end{equation}
The l.h.s. gives the change of the current noise induced
by low-frequency a.c. voltage at non-matching frequency. 
This response coefficient, and its importance,
has been recently discussed in \cite{nsethermimp},
where it has been termed "noise thermal impedance".
The authors have conjectured the relation of this coefficient to
the fourth cumulant.

It is easy to see that Eq.~(\ref{eq:UniversalRelation}) 
holds for the elastic part of the FCS even before 
integration over energies in (\ref{eq:elastic}), since
for any energy 
\begin{equation}
	\frac{f_l(1-f_r)}{f_r(1-f_l)} = e^{eV/k_\mathrm{B}T}.
\end{equation}
The corresponding proof for the inelastic contribution
can be done only after integration over energies, is 
cumbersome and has provided a good check for the validity
of expression (\ref{eq:inelastic}).

\section{Numerical Results}
\begin{figure}
	\includegraphics[width=0.95\columnwidth]{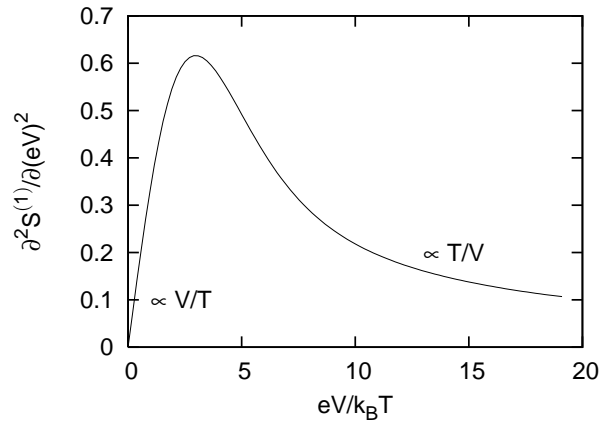}
	\caption{Second derivative with respect to voltage of the correction to the average current, $\partial^2 S^{(1)}/\partial (eV)^2\equiv\hbar k_\mathrm{B}T/[2eRG_\mathrm{Q}\sum_n T_n(1-T_n)]\partial^2 I/\partial (eV)^2$ vs
	$eV/k_\mathrm{B}T$.}
	\label{fig:d2vcur}
\end{figure}

\begin{figure}[htbp]
	\includegraphics[width=0.95\columnwidth]{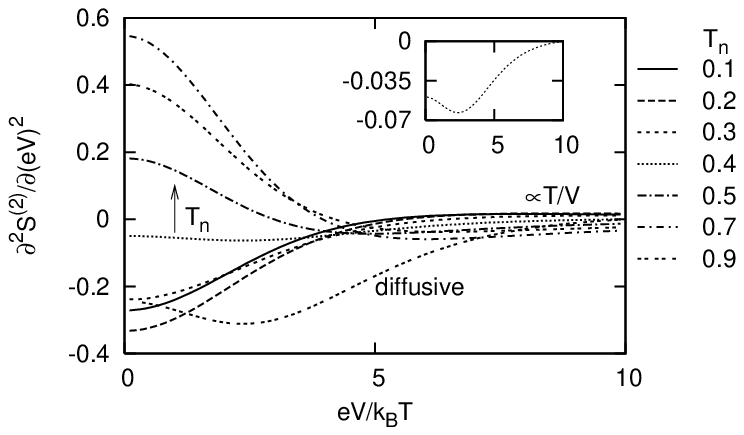}
	\caption{Second derivative with respect to voltage of the correction to the noise, $\partial^2 S^{(2)}/\partial (eV)^2\equiv\hbar k_\mathrm{B}T/[2e^2RG_\mathrm{Q}]\partial^2 I^{(2)}/\partial (eV)^2$ vs $eV/k_\mathrm{B}T$. The general tendency as a function of transmission $T_n$ is indicated by an arrow. The inset shows a zoom for $T_n=0.4$.}
	\label{fig:d2vnoise}
\end{figure}

\begin{figure}[htbp]
	\includegraphics[width=0.95\columnwidth]{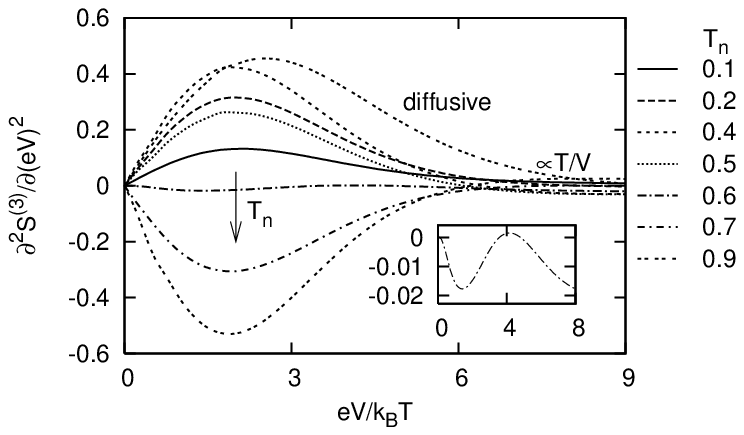}
	\caption{Second derivative with respect to voltage of the correction to the third cumulant, $\partial^2 S^{(3)}/\partial (eV)^2\equiv\hbar k_\mathrm{B}T/[2e^3RG_\mathrm{Q}]\partial^2 I^{(3)}/\partial (eV)^2$ vs $eV/k_\mathrm{B}T$. The general tendency as a function of transmission $T_n$ is indicated by an arrow. The inset shows a zoom for $T_n=0.6$.}
	\label{fig:d2vthrd}
\end{figure}

We restrict the numerical analysis to the RC-environment model discussed
above.
There are three energy scales 
in the system: voltage, temperature 
and the cut-off frequency of the environment, 
$\omega_\mathrm{c}$. 
It is especially interesting to study 
the case $\hbar\omega_\mathrm{c}\gg eV,k_\mathrm{B}T$. 
In this case it is expected 
that all cumulants show the logarithmic divergence at small 
voltages/temperature, which is well known for the conductance: 
$\delta G\sim G_\mathrm{Q}\ln\frac{\hbar\omega_\mathrm{c}}{\max(eV,k_\mathrm{B}T)}$. 
This motivated us to the following choice of
the presentation of the results: 
we plot the second derivative of the correction
to the first three  cumulants with respect to voltage  versus $eV/k_\mathrm{B}T$ (figures \ref{fig:d2vcur},\ref{fig:d2vnoise},\ref{fig:d2vthrd}). 
It is the second derivative that approaches a limit
independent of $\omega_\mathrm{c}$ upon increasing the ratios $\hbar\omega_\mathrm{c}/eV$,
$\hbar\omega_\mathrm{c}/k_\mathrm{B}T$. To illustrate the dependence on transmission and to assess differences between specific conductors (ballistic, tunnel,\ldots),
we plot the results 
for a single-channel conductor
with  transmission values ranging from $0.1$ to $0.9$. 
Another interesting reference system 
is a diffusive conductor.  
The results for a diffusive conductor 
can be obtained by averaging over 
transmission eigenvalues with 
the distribution function $\rho(T_n)=(T_n\sqrt{1-T_n})^{-1}$.
\cite{generalTn1,generalTn2}
Since all cumulants are polynomials in $T_n$ 
this is equivalent 
to replacing $T_n^m\to \sqrt{\pi}\Gamma(m)/\Gamma(m+1/2)$.

At $eV/k_\mathrm{B}T \gg 1$ the corrections are dominated by $S_\mathrm{el}$ and the emission term $S_\mathrm{in}^{(1)}(-\omega,\chi)$, since the environment does not provide any energy at $T\to 0$. The functions plotted $\propto z /V$
for  all cumulants, and are given by Eq. \ref{eq:cumulant-big-voltage}. As discussed in the paragraph below that equation, this leads to a suppression of the conductance (and any other cumulant) at small voltages which is termed zero-bias anomaly.

Let us now discuss differences between cumulants, starting with the current (fig. \ref{fig:d2vcur}). An apparent feature of the current correction is that the corrections to different conductors can be scaled to one curve by the common prefactor $\sum_nT_n(1-T_n)$, (Eq. \ref{eq:currexactlong}). The corresponding curve for a diffusive conductor can be obtained, following from averaging over transmissions, by multiplication with $2/3$ (and removing the dependence on $T_n$ in the normalization). This feature is unique for the correction to the current and independent of the choice of a specific environment. Higher order cumulants have a more involved dependence on transmission eigenvalues and environment.

In the opposite limit of large temperatures, $eV\ll k_\mathrm{B}T$, the environment provides the energy for inelastic electron transfer. Consequently the absorption term, $S^{(1)}_\mathrm{in}(\omega,\chi)$ becomes more important in equation \ref{eq:fstorder}. It is expected that in the same way as for the zero-bias anomaly the correction to the conductance is logarithmically diverging with temperature. Due to the choice of presentation, this term is not  visible in figure \ref{fig:d2vcur}. What is shown in this plot and the following is the lowest order term in the expansion in powers of $eV/k_\mathrm{B}T$. For the current (and any odd cumulant) this is a linear term, following from the symmetry with respect to inversion of voltage as explained below equation \ref{eq:inelastic}. In summary, the correction to the current shows a crossover from a temperature to voltage dominated regime at $eV\sim k_\mathrm{B}T$.

In figure \ref{fig:d2vnoise} we plotted the second derivative with respect to voltage of the correction to the noise for several single channel conductors and a diffusive conductor. 
At small voltages, all curves start with zero slope since noise is an even function of voltage. Interestingly with increasing voltage they all cross the x-axis at $eV\sim k_\mathrm{B}T$, before approaching zero. As expected, changes occur on the scale of temperature. The limit of vanishing temperature, $eV/k_\mathrm{B}T\gg 1$, namely the proportionality to $z/V$, can be discussed along the same lines as for the current. However there are several striking differences that were absent in the correction of the current.

Unlike for the current, curves for different transmission 
can not be reconciled by scaling, rather
we observe a strong dependence on the value of the transmission. This dependence is non-monotonous. However at small voltages, conductors with $T_n\lesssim 1$ have positive correction while those with $T_n\ll 1$ have a negative. This sign-change could have been conjectured since it is well known that corrections to the $m$-th cumulant are related to the unperturbed $m+1$-th cumulant. Hence, the correction to noise should be related to the third cumulant whose dependence on the transmission eigenvalues ($\propto T_n(1-T_n)(1-2T_n)$) changes sign at intermediate transmissions. 

This behavior can be interpreted by looking at the extreme cases of $T_n\ll 1$ and $T_n \lesssim 1$. It is plausible that either conductor in the absence of an environment produces little shot noise since in the first case the current is ``most of the time'' zero with only rare transfers of charges. In the second limit, electrons are transfered with probability close to one and only occasionally reflected.

The same conductors embedded in an environment however will feel a suppression of current due to fluctuations of the voltage in the node as discussed in section \ref{sec:action}. For the noise of the tunnel conductor that means, that the rare electron transfers being the source of noise, will be suppressed furthermore leading to a reduction of noise (negative correction). If $T_n \lesssim 1$, the suppressed conductance means that the quasi-constant flow of electrons will be interrupted more often, resulting in an enhancement of noise (positive correction). Consequently there will be a crossover at intermediate values of the transmission which can be seen in figure \ref{fig:d2vcur}.

Comparing the shape of the curves we observe that the maximum (absolute) value lies for most conductors at $V=0$. Remarkable exceptions are the diffusive conductor and such with transmission close to the crossover ($T_n=0.4$, inset figure \ref{fig:d2vnoise}). We might argue that the diffusive conductor inherits this feature from intermediate transmissions or rather that it is an ``interference`` effect determined by the coefficients of lowest and highest power in transmission ($T_n,T_n^3$) in the expression for the correction to the noise.

The corresponding plot of the correction to the third cumulant, which reflects the asymmetry of electron transfer, is presented in figure~\ref{fig:d2vthrd}. It shares features of both current and noise correction. Due to the different symmetry with respect to voltage inversion, the corrections to the third cumulant start at zero with linear slope. Again there is a crossover at $eV\sim k_\mathrm{B}T$ and a decay with $z/V$ at large voltages reflecting the zero bias anomaly. The dependence on $T_n$ is non-trivial which is not surprising since the expression contains four terms of different power in $T_n$, each of which can have a distinct dependence on $eV/k_\mathrm{B}T$. However ballistic/tunnel conductors at small voltages separate in the same way (albeit with opposite sign) as for the correction to noise  to the lower/upper part of the plot. Comparing curves of small $T_n$ with those for the current we recover the tunneling limit discussed in section \ref{sec:analytical}. Interestingly the correction for intermediate transmission appears to be more feature-rich than for the limiting cases. This is clearly an indication that terms of different power in $T_n$ contribute with equal weight to the interaction correction while the correction for a tunnel contact is dominated by just one term.

As the main result of our numerical analysis we note, that the corrections to cumulants strongly depend on the transmission of the contact. They can have either sign and a distinct dependence on the ratio of voltage and temperature. Both of these facilitate the experimental detection of environmental effects on transport properties. The plots in this section were obtained for an RC-environment. In principle one could obtain results for any given $z_\omega$. At least qualitatively we expect the corrections due to a (physical) environment to be similar to those presented.
\section{Conclusion}
We have studied the interaction correction to Full Counting Statistics of electron transport in a quantum contact. It was shown that the interaction can be modeled by an environmental impedance $Z_\omega$ in series with the contact. In section \ref{sec:action} we presented a formulation of the problem in terms of a non-equilibrium Keldysh action. Assuming $Z G_\mathrm{Q}\ll 1$ we proceeded perturbatively and calculated the correction to the cumulant generating function, that is, to any cumulant (Eq. \ref{eq:fstorder}).
This correction splits into three parts corresponding to elastic electron transfer and inelastic transitions with absorption/emission of energy from the environment.

We looked in detail at the structure of the interaction correction and found a re-interpretation in terms of elementary events. This provided a deeper insight into the physics involved and presented a basic check for the obtained expression. Since the full expression, Eq. \ref{eq:fstorder} is a complicated function of temperature and voltage which is not easily understood, we looked at certain limiting cases. In the limit of vanishing temperature we found a particularly simple expression for the correction to any cumulant, Eq. \ref{eq:cumulant-big-voltage}. For the opposite limit of vanishing voltage, we realized that any expression between cumulants is due to a universal relation of detailed balance for the generating function that holds irrespective of the concrete structure of the quantum contact and possible interactions, Eq. \ref{eq:UniversalRelation}.
In order to bridge those limits and to enable the experimental observation of environmental effects on electron transport in a quantum contact, we calculated numerically the correction due to an environment to the first three cumulants for arbitrary voltage, temperature and different transmission eigenvalues. We have shown that the corrections show an interesting crossover behavior from voltage to thermal noise at $eV\simeq k_\mathrm{B}T$ as well as a specific non-trivial dependence on transmission eigenvalues. The presented analytical and numerical results facilitate the measurement of the interaction correction.

\section{Acknowledgement}
This work was supported by the Foundation for Fundamental Research on Matter (FOM), The Netherlands.
\appendix
\section{Derivation of the universal relation for FCS}
In this appendix, we will present the derivation 
of the universal relation (\ref{eq:UniversalRelation})
that is valid for FCS of any conductor 
regardless its concrete realization.
A well-known example of a relation of this type is provided
by the fluctuation-dissipation theorem that relates the linear response
and Gaussian fluctuations.\cite{FDT} The same approach can be extended
to provide similar relations for non-linear frequency-dependent 
responses and non-Gaussian fluctuations \cite{Stratonovich,Chou,Wang}
using Hermicity, time-reversibility and KMS \cite{Kubo,MartinSchwinger} relations. 
These results cannot be immediately used for our purpose since
they are formulated in terms of relations between 
multi-point Keldysh Green functions rather than in terms of generating 
functions.

We apply this approach  to the most general 
generating functional of current fluctuations 
where both voltage applied $V$ and counting
field $\chi$ depend on time. The derivatives of
the functional with respect to $\chi(t)$ give averages of
current operators.
The Hamiltonian in the
presence of the voltage source can be written as
\begin{equation}
\hat{H}(t) = H_0 - \hbar\Phi(t) \hat{I}/e;\; \dot{\Phi}(t) =
eV(t)/\hbar ,
\end{equation}
$\hat{I}$ being the operator of full current in the conductor.

We make use of the interaction picture
introducing 
$\hat{I}(t) = e^{i\hat H_0t/\hbar} \hat I e^{-i\hat H_0t/\hbar}$. 
The generating function reads \cite{yuliann,kindermannnazarov,
kindermcurvoltbias} 
\begin{eqnarray}
F(\{\phi^{+}(t)\},\{\phi^{-}(t)\}) = \langle 
\hat{U}^{\dagger}(\{\phi^{-}(t)\})\hat{U}(\{\phi^{+}(t)\}) \rangle,
\label{Zaverage}\\
\hat{U}(\{\phi^{+}(t)\}) = \overrightarrow{T} \exp \left( i \int dt 
\;\phi^{+}(t) \hat{I}(t)/e \right) , \\
\hat{U^{\dagger}}(\{\phi^{-}(t)\}) = \overleftarrow{T} \exp \left( -i \int dt 
\;\phi^{-}(t) \hat{I}(t)/e \right) , \\
\langle \dots \rangle = {\rm Tr} \left( \dots \hat{\rho} \right);\; 
\hat \rho = e^{-\hat{H}/k_\mathrm{B}T}/{\rm Tr}\left(e^{-\hat{H}/k_\mathrm{B}T}\right),
\end{eqnarray}
where $\overrightarrow{T}(\overleftarrow{T})$ stands for (anti)time-ordering of the operators and $\phi^{\pm}(t) = \varphi(t) \pm \chi(t)/2$. 
This expression is formally equivalent to the
generating functional for multi-point Keldysh Green
functions used in \cite{Chou,Wang}. The only difference is
that the Green functions generated are those of current
operators.

We shall assume time-reversibility of the Hamiltonian.
Since in this case
\begin{equation}
\hat{H}^{T} = \hat{H};\; \hat{I}^T = -\hat{I};\; \hat{I}^T(t) = - \hat{I}(-t)\\
\end{equation}
we observe the following transposing rule for $S$-operators:
\begin{equation}
\left(\hat{U}(\{\phi(t)\})\right)^T = \hat{U} (\{-\phi(-t)\})   
\end{equation}
Transposing operators in the average (\ref{Zaverage}), we obtain
$$
F(\{\phi^{+}(t)\},\{\phi^{-}(t)\}) = \langle \hat{U}(\{-\phi^{+}(-t)\})
\hat{U}^{\dagger}(\{-\phi^{-}(-t)\}) 
\rangle . 
$$
In comparison with (\ref{Zaverage}), $U,U^{\dagger}$ are interchanged. 
We want them
back to their original positions.
The way to do this is to make use of KMS relations: For any operator $\hat{A}$
\begin{equation}
	\hat{\rho} \hat{A}(t) = \hat{A}(t + i\hbar/ k_\mathrm{B}T) \hat{\rho}. 
\end{equation}
We do this commutation with all operators comprising $U^{\dagger}$ to obtain
\begin{eqnarray}
&\hat{\rho} \hat{U}(\{-\phi^+(-t)\}) = & \nonumber\\
 =&\overrightarrow{T}  \exp\left(- i \int dt 
 \phi^{+}(-t)  \hat{I}(t+i\hbar/k_\mathrm{B}T)/e \right)\hat{\rho}& .
\end{eqnarray}
We shift now the time argument of $\phi^{+}$ by $i \hbar/k_\mathrm{B}T$ to obtain
\begin{eqnarray*}
\overrightarrow{T}  \exp \left( -i \int dt 
\phi^{+}(-t) \hat{I}(t+i\hbar/k_\mathrm{B}T)/e \right) = \\
\hat{U}( \{-\phi^{+}(-t+i\hbar/k_\mathrm{B}T) \}). 
\end{eqnarray*}
This step looks rather heuristic. Since nothing is assumed
concerning the analytical properties of $\phi$ as a function of
complex time, the complex shift may be ambiguous.  
However we note that we are mainly interested in $\phi^{\pm}(t)$
that change at time scales much bigger than $\hbar/k_\mathrm{B}T$: for those, 
we expect no ambiguity. 

Finally, we cycle operators under the sign of trace to obtain
\begin{eqnarray}
F(\{\phi^{+}(t)\},\{\phi^{-}(t)\}) = \nonumber \\
F(\{-\phi^{+}(-t+i\hbar/k_\mathrm{B}T)\},\{-\phi^{-}(-t)\}).
\end{eqnarray}

For quasi-stationary $V,\chi$ we substitute $\phi^{\pm} = eVt/\hbar  \pm \chi/2$
and neglect the dependence on time-independent phase to arrive to
\begin{equation}
	F(V,\chi) = F(V, -\chi + i eV/k_\mathrm{B}T) 
\end{equation}
which is the universal relation to prove.

\bibliography{fisica}
\end{document}